\documentstyle[12pt]{article}
\newcommand{\lsim}{\stackrel{\mbox{\raisebox{-0.1ex}{\scriptsize $<$}}}
{\mbox{\raisebox{-0.5ex}{\scriptsize $\sim$}}}}
\newlength{\myleftmargin}
\newlength{\paperwidth}
\begin{document}
\renewcommand{\thefootnote}{\fnsymbol{footnote}}
\begin{flushright}
KOBE--FHD--94--08~\\
October~~~20~~~1994
\end{flushright}
\begin{center}
{\large \bf The Shape of Polarized Gluon Distributions}\\
\vspace{3.5em}
T. Morii\footnote[2]{E--mail~~~~morii@jpnyitp.bitnet}\\
\vspace{0.8em}
{\it Faculty of Human Development, Division of}\\
{\it Sciences for Natural Environment}\\
{\it and}\\
{\it Graduate School of Science and Technology,}\\
{\it Kobe University, Nada, Kobe 657, Japan}\\
\vspace{1.5em}
S. Tanaka\\
\vspace{0.8em}
{\it Faculty of Human Development, Division of}\\
{\it Sciences for Natural Environment,}\\
{\it Kobe University, Nada, Kobe 657, Japan}\\
\vspace{1em}
and\\
\vspace{1em}
T. Yamanishi\footnote[8]{E--mail~~~~yamanisi@natura.h.kobe--u.ac.jp}\\
\vspace{0.8em}
{\it Graduate School of Science and Technology,}\\
{\it Kobe University, Nada, Kobe 657, Japan}\\
\vspace{2.5em}
{\bf Abstract}
\end{center}

\vspace{1em}

\baselineskip=24pt

The recent high precision SMC data on polarized $\mu p$ scatterings
have again confirmed that very little of the proton spin is carried
by quarks. To unravel the mystery of the proton spin structure,
it is quite important to know the behavior of the polarized gluon distribution.
By using the positivity condition of distribution functions together with the
unpolarized and polarized experimental data, we restrict the $x$ dependence of
the polarized gluon distribution.

\vfill\eject

Recently, the SMC group\cite{SMC94} at CERN measured the spin--dependent
proton structure function $g_1^p(x)$ more precisely and to the smaller $x$
region up to $x=0.006$ than the previous measurements carried out by the
EMC\cite{EMC}. The experiment indicates that the first moment of
$g_1^p(x)$ increases about $10\%$ compared to the EMC result, and yet that
value is still far from the value predicted by the nonrelativistic quark model
and the one from the Ellis--Jaffe sum rule\cite{Ellis}. By combining these SMC
data with the experimental data of the neutron $\beta$--decays and hyperon
$\beta$--decays, the polarized strange quark density in the proton is derived
as follows:
\begin{equation}
\Delta s = -0.12\pm 0.04\pm 0.04~.
\label{eqn:Ds}
\end{equation}
On the other hand, Preparata, Ratcliffe and Soffer have shown that a bound on
the value of $\Delta s$ can be obtained by requiring the positivity of
distribution functions and assuming the reasonable behavior of the
unpolarized s--quark distribution $s(x)$\cite{Preparata91}.
Quite contrary to the SMC result, they got
\begin{equation}
|\Delta s| \leq 0.021\pm 0.001~,
\label{eqn:Ds-bound1}
\end{equation}
by using the $s(x)$ derived from the $\nu$N
deep--inelastic scattering experiments\cite{CCFR}. Furthermore, similar
results were
obtained by Preparata and Soffer who indicated the following bound on the
polarized s--quark density\cite{Preparata88}:
\begin{equation}
|\Delta s| \leq 0.05^{+0.02}_{-0.05}~,
\label{eqn:Ds-bound2}
\end{equation}
using CDHS\cite{CDHS} and WA25 data\cite{WA25}. At first sight, these
bounds seem to be contradictory to the SMC data
of eq.(\ref{eqn:Ds}). There might be, however, a compromising solution. If the
gluons contribute to the proton spin through the U$_A$(1)
anomaly\cite{anomaly}, the left--hand side of eq.(\ref{eqn:Ds}) should be
modified as
\begin{equation}
\Delta s \rightarrow \Delta s-\frac{\alpha_S}{2\pi}\Delta G~,
\label{eqn:anomaly}
\end{equation}
where $\Delta G$ denotes the polarization of gluons. Then the bound of
$|\Delta s|$ given by
(\ref{eqn:Ds-bound1}) and (\ref{eqn:Ds-bound2}) turns out to be
consistent with the SMC data of eq.(\ref{eqn:Ds}) by taking rather
large $\Delta G$ ($\simeq 5 - 6$). Namely $\Delta s$ remains small with the
cost of large $\Delta G$. Moreover, with this prescription quarks
are to carry most of the proton spin and hence one can realize naturally
the quark--parton picture. Therefore it is very important to
know the magnitude of $\Delta G$ and the $x$ dependence of the polarized gluon
distribution $\delta G(x)$, where $\Delta G=\int^1_0 \delta G(x)dx$.
So far there have been
some interesting studies on the polarized gluon. In literature, various types
of the polarized gluon distribution functions have been proposed: some of them
have large $\Delta G$ ($\simeq 5 - 6$)\cite{Altarelli89,Cheng90,Rams91}
and others have small $\Delta G$
($\lsim 2 - 3$)\cite{Cheng90,Rams91,Brod94,Gehrmann}.
The E581/704 collaboration\cite{E581} measured the two--spin asymmetries
$A^{\pi^0}_{LL}(\stackrel{\scriptscriptstyle(-)}{p}\stackrel{}{p})$
for $\pi^0$ productions in polarized
$\stackrel{\scriptscriptstyle(-)}{p}\stackrel{}{p}$ collisions and concluded,
by comparing the measured asymmetries
$A^{\pi^0}_{LL}(\stackrel{\scriptscriptstyle(-)}{p}\stackrel{}{p})$
with the theoretical predictions by Ramsey et al\cite{Rams91},
that the large $\Delta G$ should be ruled out.
However, some people\cite{Weber} have pointed out that the calculations
significantly depend on the shape of polarized gluon distribution functions
and hence the large $\Delta G$ is not necessarily ruled out but the shape
of $\delta G(x)$ is strongly constrained by the E581/704 data.

In this work, we study the $x$ dependence of the polarized gluon distribution
$\delta G(x)$.
In the previous papers\cite{Kobayakawa92,Morii94}, we have proposed a simple
model of polarized distributions of quarks and gluons
which reproduce the EMC experimental data well. In this model
$\Delta s$ was determined to be rather small such as
$0.019$, which was consistent with the bound of (\ref{eqn:Ds-bound1}) and
(\ref{eqn:Ds-bound2}). As for the magnitude of $\Delta G$,
we can fix its value to be $5.32$ from the experimental data of the integral
value of $g_1^p(x)$. However, as for the $x$ dependence of $\delta G(x)$,
nobody knows the exact form of it at present: there remains a number of
unknown factors in $\delta G(x)$, which cannot be calculated perturbatively.
Here by taking account of the plausible behavior of the distribution
$\delta G(x)$ near $x\approx 0$ and $x\approx 1$, we assume
\begin{eqnarray}
\delta G(x) &=& G_+(x) - G_-(x) \nonumber\\
            &=& B~x^{\gamma}~(1-x)^p~(1+C~x)~,
\label{eqn:delta G}
\end{eqnarray}
where $G_+(x)$ and $G_-(x)$ are the gluon distributions with helicity parallel
and antiparallel to the proton helicity, respectively. We further assume for
simplicity $G_+(x)\approx G_-(x)$ at large $x$ and take $C=0$. Then there
remain two parameters, $\gamma$ and $p$. $B$ is determined from the
normalization, $\Delta G=5.32$. We are interested in the behavior of
$\delta G(x)$
under the condition of large $\Delta G$ ($=5.32$) and study the allowable
region of $\gamma$ and $p$. In order to implement this,
we require the positivity condition of distribution functions and utilize
the recent results of several polarization experiments.

As a preliminary, to examine the behavior of $\delta G(x)$ in
eq.(\ref{eqn:delta G}) for various values of $\gamma$ and $p$, we vary
$\gamma$ from $-0.9$ to $0.3$ at intervals of $0.3$ while we choose $p$
independently as $5$, $10$, $15$, $17$ and $20$. The results are presented in
Fig.1. One can see from this figure that if one takes $\gamma$ smaller, the
peak of the distribution is shifted to smaller $x$ and if one takes smaller
$p$, the distribution has a broader shoulder.

Now, let us get into the discussion on the restriction of the $x$ dependence
of $\delta G(x)$.

(i)~~First, we consider the positivity condition of distribution functions to
restrict $\gamma$ and $p$. As for the unpolarized gluon distribution $G(x)$,
we assume
\begin{eqnarray}
G(x) &=& G_+(x) + G_-(x)\nonumber\\
     &=&\frac{A}{x^{\alpha}}~(1-x)^k
\label{eqn:G(x)}
\end{eqnarray}
like in the case of eq.(\ref{eqn:delta G}). Since $G_+(x)$ and $G_-(x)$
are both positive, we obtain from eqs.(\ref{eqn:delta G}) and (\ref{eqn:G(x)})
\begin{equation}
|~B~x^{\gamma}~(1-x)^p~| \leq \frac{A}{x^{\alpha}}~(1-x)^{k}~.
\label{eqn:posit1}
\end{equation}
{}From eq.(\ref{eqn:posit1}) we get
\begin{equation}
|~B~| \leq \frac{A}{x^{\alpha +\gamma}}~(1-x)^{k-p}~,
\label{eqn:posit2}
\end{equation}
and
\begin{equation}
|~\Delta G~| \leq \frac{\Gamma (\gamma +1)~\Gamma (p+1)~\Gamma (k+3-\alpha)~
                 (\alpha +\gamma+p-k)^{\alpha +\gamma+p-k}}
          {\Gamma (\gamma +p+2)~\Gamma (k+1)~\Gamma (2-\alpha)~
          (\alpha +\gamma)^{\alpha +\gamma}(p-k)^{p-k}}\int^1_0 xG(x)dx~,
\label{eqn:bound}
\end{equation}
To restrict the region of $\gamma$ and $p$ from
this inequality (\ref{eqn:bound}) with
$\Delta G=5.32$, we need to know the value of $\alpha$ and $k$ in $G(x)$ and
the intergral value of $xG(x)$ as well. As for the $x$ dependence of $G(x)$,
using  experimental data of J/$\psi$ productions for unpolarized muon--nucleon
scatterings\cite{NMC91,EMC92}, we have two possible types of parameterization
of $G(x)$ at $Q^2\simeq M_{J/\psi}^2$ GeV$^2$,
\begin{eqnarray}
Type~A~~~~~~~~~G(x) &=& 3.35~\frac{1}{x}~(1-x)^{5.7}~,
\label{eqn:type-a}\\
Type~B~~~~~~~~~G(x) &=& 2.36~\frac{1}{x^{1.08}}~(1-x)^{4.65}~.
\label{eqn:type-b}
\end{eqnarray}
For Type A, $\alpha$ is taken to be $1$ by considering the ordinary Pomeron P,
and parametrized so as to fit the data. On the other hand, $\alpha$ is chosen
to be $1.08$ in Type B which is recently derived from the analysis of the
experimental data of the total cross section\cite{Donnachie}. The graphs of
these two distributions are given in Fig.2,
where the intergral values of $xG(x)$ in eqs.(\ref{eqn:type-a}) and
(\ref{eqn:type-b}) are both normalized to $0.5$ in conformity to the
experimental data. Inserting these functions into inequality
(\ref{eqn:bound}) with $\Delta G=5.32$, the allowed regions of $\gamma$ and
$p$ are obtained.
We have examined (\ref{eqn:bound}) for various combinations of $\gamma$ and
$p$, and the results are given in Table 1 and Fig.3.
In Fig.3, the region below solid
or dashed lines is excluded by (\ref{eqn:bound}).
{}From this analysis, we conclude that a wide region of $\gamma$ and $p$ which
satisfies the SMC data and the positivity condition simultaneously, is
allowable with respect to the polarized gluon distribution with large
$\Delta G$ ($=5.32$).

\vspace{2em}

(ii)~~Second, to restrict further the allowable region of $\gamma$ and $p$, we
compare our model calculations with the two--spin asymmetries
$A^{\pi^0}_{LL}(\stackrel{\scriptscriptstyle(-)}{p}\stackrel{}{p})$ for
inclusive $\pi^0$--productions measured by E581/704 Collaboration using
polarized proton (antiproton) beams and polarized proton targets\cite{E581}.
Taking $\delta G(x)$ with the combination of $(\gamma, p)$ which is
allowed by the criterion of positivity, we calculate numerically
$A^{\pi^0}_{LL}(\stackrel{\scriptscriptstyle(-)}{p}\stackrel{}{p})$, where the
polarized quark distributions $\delta q_i(x)$, which are necessary for
the calculation of cross sections for some of subprocesses, are taken from
ref.\cite{Kobayakawa92}. The results are given in Fig.4.
{}From this figure, some combinations of $\gamma$ and $p$ are
excluded. Surviving combinations of $(\gamma, p)$ are shown in Table 2.
Comparing the calculations with the experimental data, we have found that
$x\delta G(x)$ must have a peak at a smaller $x$ than
$0.05$ and has to decrease very rapidly with increasing $x$. In short,
the experimental data are reproduced well when $\gamma$ is small and $p$ is
large, though it is rather difficult to say which one is the best fitting.

\vspace{2em}

(iii)~~Finally, we look into the spin--dependent structure function of proton
$g_1^p(x)$\cite{SMC94} and that of deuteron $g_1^d(x)$\cite{SMC93}.
The merit of considering these parameters is that $g_1^p(x)$ and $g_1^d(x)$
do not include undetermined fragmentation functions which were included in
$A^{\pi^0}_{LL}(\stackrel{\scriptscriptstyle(-)}{p}\stackrel{}{p})$.
Accordingly they are more sensitive to the behavior of $\delta G(x)$.
For this case $\gamma$ might be bounded below, while this is not
the case for the former two cases. For example, for $\gamma=-0.9$ the
calculated values of $xg_1^p(x)$ seem to deviate from the data for small $x$
regions, $x<0.005$. The new SMC data\cite{SMC94} show a tendency for
$g_1^p(x)$ to increase for small $x$, $x<0.01$, while the calculated values
with $\gamma=-0.9$ keep decreasing for such a small $x$ region. It is
expected that if $\gamma$ gets smaller, the discrepancy of $g_1^p(x)$ between
the calculated values and the experimental data would become larger.
In addition, for $g_1^d(x)$ the calculation with $\gamma=-0.9$ does
not fit well to the data for $0.01<x<0.05$. The result of calculation using
our $\delta q_i(x)$ and $\delta G(x)$ with $(\gamma, p)$ surviving
the criteria of cases (i) and (ii) is shown in Fig.5 and Table 3.

In summary, in the models with large $\Delta G$ ($=5.32$), we have studied
the shape of the polarized gluon distribution. By using the positivity
condition of distribution functions together with the experimental data on
the two--spin asymmetries
$A^{\pi^0}_{LL}(\stackrel{\scriptscriptstyle(-)}{p}\stackrel{}{p})$
and the spin--dependent structure functions of $g_1^p(x)$ and $g_1^d(x)$,
we have restricted the $x$ dependence of $\delta G(x)$ as given in
eq.(\ref{eqn:delta G}). As for the magnitude of $\gamma$,
$-0.6\lsim\gamma\lsim -0.3$ seems favorable in our analysis, and with respect
to $p$ we obtain the bound that $p$ should be larger than 15.
In other words, if $\gamma$ and $p$ are fixed in this region, for example,
as $\gamma=-0.6$ and $p=17$, one can reproduce
all existing data quite successfully. Needless to say, the $\Delta s$ of
eq.(\ref{eqn:Ds}) can be reconciled with the bound of (\ref{eqn:Ds-bound1}) or
(\ref{eqn:Ds-bound2}) with large $\Delta G$ ($=5.32$).
However, at present we do not know the theoretical ground on the origin of
these values of $\gamma$ and $p$: in the Regge terminology, the value of
$\gamma$ restricted above happens to be closer to the one for unpolarized
valence quark distributions rather than for unpolarized gluon
distributions\cite{Regge}, and $p$ seems to be inconsistent with the
prediction of counting rules\cite{count}. To understand the origin of such
$\gamma$ and $p$ is out of scope in this work and needs further
investigations. Furthermore, if $\Delta G$ is so large
($\simeq 5 - 6$), we are to have an approximate relation
$\langle L_Z\rangle_{q+G}\simeq -\Delta G$ from the proton spin sum rule,
$\frac{1}{2}=\frac{1}{2}\Delta\Sigma+\Delta G+\langle L_Z\rangle_{q+G}$,
where $\frac{1}{2}\Delta\Sigma$ represents the sum of the spin carried by
quarks. Unfortunately, nobody knows the underlying physics of it. These are
still problems to be solved even though the idea of the U$_A$(1) anomaly is
attractive.

It is informative to comment on another approach which has mentioned to this
problem.
Recently Brodsky, Burkardt and Schmidt (BBS)\cite{Brod94} have proposed an
interesting model of the polarized gluon distribution which incorporates
color coherence and counting rule at small and large $x$.
At $x\approx 0$, the color coherence argument gives $\delta G(x)/G(x)\approx
\frac{x}{3}\langle\frac{1}{y}\rangle$ with $\langle\frac{1}{y}\rangle\simeq 3$,
where $\langle\frac{1}{y}\rangle$ presents the first inverse moment of the
quark light--cone momentum fraction distributions in the lowest Fock state of
the proton, and leads to a relation $\gamma=-\alpha +1$\cite{Brod94,Gehrmann}.
Then, contrary to our result, $-0.6\lsim\gamma\lsim -0.3$, they have taken
$\gamma=0$ by choosing $\alpha=1$ which is an ordinary Pomeron intercept value.
In terms of Regge theory, $\gamma =0$ can be interpreted as follows:
$\delta G(x)$ at
$x\approx 0$ is governed by the $A_1$ trajectory. Although the integrated
value of $\delta G(x)$ in the BBS model is small such as $\Delta G=0.45$,
the model is successful in explaining the EMC data $g_1^p(x)$, $g_1^n(x)$ and
$g_1^d(x)$. In addition, we calculated
$A^{\pi^0}_{LL}(\stackrel{\scriptscriptstyle(-)}{p}\stackrel{}{p})$
by using the BBS model and found that the model could reproduce
$A^{\pi^0}_{LL}(\bar pp)$ while the predicted value of $A^{\pi^0}_{LL}(pp)$
slightly deviated from the data\cite{Morii94}. The BBS model which has
small $\Delta G$ ($=0.45$) seems to be an alternative to our model which
has large $\Delta G$ ($=5.32$), though in the BBS model the apparent
inconsistency between $\Delta s$ of eq.(\ref{eqn:Ds}) and the bound of
(\ref{eqn:Ds-bound1}) or (\ref{eqn:Ds-bound2}) remains to be unsolved.

It is very important to know the behavior of $\delta G(x)$ and $\delta s(x)$
in order to understand the spin structure of a proton. However the polarization
experiments are still in their beginning and the form of these functions
is not yet clear disappointingly. We hope they will be determined in the
forthcoming experiments.

\vspace{2em}

\begin{center}
{\bf Acknowledgments}
\end{center}
The authors are grateful to N. I. Kochelev and G. Ramsey for their important
comments on the behavior of polarized gluons at very small $x$ regions.
This work was supported in part by Grant--in--Aid for Scientific Research
No.20031370 from the Ministry of Education, Science and Culture, Japan.

\vfill\eject

\vfill\eject

\begin{center}
{\large \bf Figure captions}
\end{center}
\begin{description}
\item[Fig. 1:] The $x$ dependence of the spin--dependent gluon distribution
function $x\delta G(x, Q^2)$ at $Q^2=10$GeV$^2$ for various $p$ ($=5 - 20$)
with (a) $\gamma=0.3$, (b) $\gamma=-0.6$ and (c) $\gamma=-0.9$.

\parskip 1em

\item[Fig. 2:] The parametrization of the gluon distribution
functions $xG(x, Q^2)$ at $Q^2\approx 10$GeV$^2$. The solid (dashed) line
denotes Type A (B). The data of opened (closed) circles are taken from
\cite{NMC91} (\cite{EMC92}).

\parskip 1em

\item[Fig. 3:] The allowed region by (\ref{eqn:bound}) for $\gamma$ and $p$.
The solid (dashed) line corresponds to Type A (B). The region below the
lines are excluded.

\parskip 1em

\item[Fig. 4:] The produced $\pi^0$ transverse momenta $p_T$ dependence of
$A^{\pi^0}_{LL}(\stackrel{\scriptscriptstyle(-)}{p}\stackrel{}{p})$ for
various $p$ ($=5 - 20$) with (a) $\gamma=0$, (b) $\gamma=-0.6$ and (c)
$\gamma=-0.9$. Data are taken from \cite{E581}.

\parskip 1em

\item[Fig. 5:] The $x$ dependence of $xg_1^p(x)$ and $xg_1^d(x)$
for various $p$ ($=5 - 20$) with (a) $\gamma=0$, (b) $\gamma=-0.6$ and (c)
$\gamma=-0.9$. The data of $xg_1^p(x)$ ($xg_1^d(x)$) are taken from
\cite{SMC94,EMC} (\cite{SMC93}).
\end{description}
\end{document}